\documentclass{naturemodified}
\usepackage{graphicx}
\usepackage{amsmath}
\usepackage{amssymb}

\title{Strong cooperative coupling of pressure-induced magnetic order and nematicity in FeSe}

\author{K. Kothapalli$^{1,2,*}$, A. E. B\"ohmer$^{1,*}$, W. T. Jayasekara$^{1,2}$, B. G. Ueland$^{1,2}$, P. Das$^{1,2}$, A. Sapkota$^{1,2}$, V. Taufour$^{1}$, Y. Xiao$^3$, E. E. Alp$^3$, S. L. Bud'ko$^{1,2}$, P. C. Canfield$^{1,2}$, A. Kreyssig$^{1,2}$ \& A. I. Goldman$^{1,2}$}

\begin{document}

\maketitle

\begin{affiliations}
\item{Ames Laboratory, U.S. DOE, Iowa State University, Ames, Iowa 50011, USA}
\item{Department of Physics and Astronomy, Iowa State University, Ames, Iowa 50011, USA}
\item{Advanced Photon Source, Argonne National Laboratory, Argonne, Illinois 60439, USA}
\end{affiliations}

\begin{abstract}
  A hallmark of the iron-based superconductors is the strong coupling between magnetic, structural and electronic degrees of freedom\cite{Johnston2010,Stewart2011}.  However, a universal picture of the normal state properties of these compounds has been confounded by recent investigations of FeSe where the nematic (structural) and magnetic transitions appear to be decoupled.  Here, using synchrotron-based high-energy x-ray diffraction and time-domain M\"{o}ssbauer spectroscopy, we show that nematicity and magnetism in FeSe under applied pressure are indeed strongly coupled.  Distinct structural and magnetic transitions are observed for pressures, 1.0~GPa~$\lesssim$~$p$~$\lesssim$~1.7 GPa, which merge into a single first-order phase line for $p$~$\gtrsim$~1.7~GPa, reminiscent of what has been observed, both experimentally and theoretically, for the evolution of these transitions in the prototypical doped system, Ba(Fe$_{1-x}$Co$_x$)$_2$As$_2$\cite{Kim2011}.  Our results support a spin-driven mechanism for nematic order in FeSe and provide an important step towards a universal description of the normal state properties of the iron-based superconductors.
\end{abstract}

Unconventional superconductivity (SC) is generally believed to be mediated by fluctuations associated with ordering (magnetic, orbital or charge) found in the normal state of the material\cite{Johnston2010,Stewart2011}. For the high-temperature iron-based superconductors, a leading candidate is the ubiquitous stripe-like antiferromagnetic (AFM) fluctuations associated with the magnetic ordering in the normal state.  In some scenarios, these same spin fluctuations also lead to "nematic" electronic order, identified by a tetragonal-to-orthorhombic (T-OR) structural transition\cite{Nandi2010,Fernandes2014}. However, the magnetic origin of nematicity has been recently challenged by the unusual properties of FeSe\cite{Watson2015,Boehmer2015,Baek2015,Massat2016,Wang2015III,Shamoto2015,Yu2015}.

For most iron-based superconductors, including $A$Fe$_2$As$_2$ ($A$ = Ca, Sr, Ba) and $R$FeAsO ($R$ = rare earth) and NaFeAs, stripe-like AFM order and nematic order, in the form of a T-OR transition, occur either simultaneously or in relatively close proximity in temperature, and both can be tuned by doping or applied pressure. At ambient pressure, FeSe undergoes a T-OR transition at $T_\mathrm{s}=90$ K, but no ordered AFM state has been observed\cite{McQueen2009}. Evidence both for and against the presence of magnetic order in FeSe under applied pressure has been reported\cite{Medvedev2009,Imai2009,Bendele2010,Bendele2012}, and when magnetic order has been found, the magnetic transition temperature, $T_{\rm{m}}$, was reported to \emph{increase} with pressure whereas the structural transition was suppressed\cite{Terashima2015,Sun2015}. Furthermore, superconductivity in FeSe shows a four-fold increase to a transition temperature of $T_{\rm{c}}=37$ K under pressure\cite{Medvedev2009}, accompanying the increase of $T_{\rm{m}}$\cite{Miyoshi2014,Terashima2015,Sun2015}. Taken together, the absence of magnetic order in the presence of nematic order at ambient pressure, and the opposing evolutions of magnetic and nematic order with pressure, cast doubt on the notion of a cooperative relation between magnetism and structure as the origin of the nematic order in FeSe\cite{Watson2015,Boehmer2015,Baek2015,Massat2016,Wang2015III,Shamoto2015,Yu2015}.

Electrical resistivity measurements under applied pressure provide valuable information regarding the existence of phase transitions in the $p$-$T$ phase space, but microscopic measurements under pressure are required to fully elucidate the underlying nature of the transitions, the associated order parameters, and the relationship between the magnetism and structure of FeSe. Although direct microscopic information concerning the magnetic structure is best obtained using  neutron diffraction, several previous measurements have been unsuccessful in detecting the expected small ordered magnetic moment \cite{Bendele2012,Sun2015}. Therefore, we have employed both high-energy x-ray diffraction (HE-XRD), on Station 6-ID-D, and time-domain M\"{o}ssbauer spectroscopy through nuclear forward scattering (NFS), on Station 16-ID-D, at the Advanced Photon Source to study the structural and magnetic transitions in FeSe under pressure. An important feature of our measurements is that they were each performed on FeSe single crystals prepared using KCl/AlCl$_3$ chemical vapor transport\cite{Boehmer2013}, which were thoroughly characterized in previous studies\cite{Kaluarachchi2016,Tanatar2016}. For such high-quality single-crystals, both $T_{\rm{s}}$ and $T_{\rm{m}}$ were observed in resistivity measurements, which facilitated a mapping of the phase diagram\cite{Terashima2015,Kaluarachchi2016,Sun2015}. The batch used for the HE-XRD experiment was characterized with DC magnetization and electrical resistivity measurements, and yielded sharp superconducting transitions at $T_{\rm{c}}=8.7-8.8$ K at ambient pressure. Samples for the NFS experiment were prepared similarly using 94\% isotopically pure $^{57}$Fe, and had $T_{\rm{c}}$ values between 8.5~-~8.8~K.

The essential results of our study are summarized in Fig. 1. Figure 1a shows that FeSe undergoes a continuous T-OR transition at ambient pressure at $T_{\rm{s}}=90$ K.  This is observed as a splitting of the ($HH0$) Bragg peaks related to the in-plane orthorhombic lattice parameters $a_{\rm{OR}}$ and $b_{\rm{OR}}$. At an applied pressure of 1.5 GPa (Fig. 1b), this structural transition is suppressed to $T_{\rm{s}}=32$ K.  However, as the temperature is further lowered there is a subsequent abrupt increase of the splitting at 19 K. When the pressure is increased to 1.7 GPa only the discontinuous T-OR transition remains, and $T_{\rm{s}}$ increases to 35 K at 3.1 GPa (Fig. 1d). In order to correlate the features observed in the HE-XRD measurements with the evolution of the magnetism in FeSe, Figs. 1e and f show the NFS (time-domain M\"{o}ssbauer spectra) at two different pressures, 2.5 GPa and 4.0 GPa. The small hyperfine field associated with magnetic ordering in FeSe is quite challenging for conventional M\"{o}ssbauer spectroscopy\cite{Medvedev2009}, but is readily observed in the data from our NSF measurements. The distinctive feature in these spectra is the minimum at intermediate delay times, which is determined by a convolution of the magnetic hyperfine field, quadrupolar splitting, and effective sample thickness (see Methods Section). Our fits to these spectra find that the shift in the minima at approximately 30~K ($p$~=~2.5~GPa) and 40~K ($p$~=~4~GPa) is attributable to the presence of a hyperfine field at these temperatures and pressures, which confirms the onset of magnetic order in FeSe under pressure.

The order parameters determined from fits to the HE-XRD data, and the evolution of the hyperfine field are shown in Fig. 2 and summarized in the $p$-$T$ diagram of Fig. 3, and provide insight into the relationship between structure and magnetism in FeSe under pressure.   At ambient pressure, the orthorhombic distortion $\delta(T)=(a_{\rm{OR}}-b_{\rm{OR}})/(a_{\rm{OR}}+b_{\rm{OR}})$ displays a smooth temperature dependence characteristic of a second-order phase transition (Fig. 2a).  Starting from approximately 1 GPa, however, we observe a small discontinuous increase in $\delta$ at approximately 19 K, which is very distinct at $p$~=~1.5 GPa (see the inset of Fig. 2a) and higher applied pressures, as shown in Fig. 2c. Similar measurements of the canonical BaFe$_2$As$_2$-type iron-based superconductors clearly established that this discontinuous increase in the orthorhombicity was associated with strong magnetoelastic coupling and a transition to stripe-like magnetic order\cite{Kim2011,Pajerowski2013}.  In Fig. 2b we show the evolution of the hyperfine field, $H_{\rm{hf}}$, extracted from fits to the NFS data measured at applied pressures of 2.5 and 4.0 GPa (Fig. 1e and 1f). The relatively small saturated values of approximately 2.5 and 3.0~T, respectively, correspond to a small ordered moment on the order of 0.2~$\mu_{\rm{B}}$/Fe, consistent with the estimates from previous zero-field muon spin-resonance ($\mu$SR) experiments\cite{Bendele2012} and the difficulty in detecting a magnetic signal in earlier conventional M\"{o}ssbauer measurements\cite{Medvedev2009}.  We also find that $T_{\rm{N}}$ and the saturation value of $H_{\rm{hf}}$ increases with increasing pressure, consistent with the previous $\mu$SR studies\cite{Bendele2010,Bendele2012}.  Although the density of data points is rather low, Fig. 2b suggests that the NFS spectra at 4.0 GPa is consistent with a strong discontinuous transition, whereas the magnetic transition at 2.5 GPa could be described as weakly first-order.  

Perhaps most importantly, the HE-XRD data show that the T-OR transition does not vanish above 1.7 GPa where $T_{\rm{s}}$, as inferred from resistivity measurements, would extrapolate to zero\cite{Terashima2015,Wang2015III}. Rather, the abrupt onset of the structural transition and the significant coexistence range of the tetragonal and orthorhombic phases (inset of Fig. 2c) demonstrate that the structural and magnetic transitions merge to a simultaneous first-order magneto-structural transition as shown in Fig. 3 and also found, for example, in the CaFe$_2$As$_2$ parent compound\cite{Ni2008II}.  Further, our HE-XRD measurements show that the magnetic order in FeSe breaks the tetragonal symmetry of the lattice in the same manner as the ubiquitous stripe-type magnetic order in the other iron-based materials.  To our knowledge, no other AFM order observed in iron-based superconductors or magnetic structures proposed for FeSe\cite{Glasbrenner2015}, break the tetragonal symmetry in this specific way.

 The HE-XRD and NFS results for FeSe demonstrate that all features concerning the normal state structure and magnetism, and their coupling, are very similar to other iron-based superconductors, and differ only in the details of the temperature and pressure dependencies of the transitions.  These results lend support to a spin-driven mechanism for nematic ordering in FeSe.  In Figure 4, we compare the schematic phase diagrams of BaFe$_2$As$_2$ and FeSe with trends derived from recent theoretical studies of spin-driven nematicity\cite{Fernandes2014}. Figure 4a sketches the sequence of structural and magnetic phase transitions for BaFe$_2$As$_2$ as a function of doping.  For Co substitution for Fe in BaFe$_2$As$_2$ ("electron doping"), the magnetic and structural transitions separate in temperature and both $T_{\rm{s}}$ and $T_{\rm{m}}$ decrease with increased substitution\cite{Kim2011}.    On the other hand, for "hole doping", the magnetic and structural transitions are concomitant and both transitions decrease in temperature with increased doping.  The $p-T$ diagram for FeSe in Fig. 4b mirrors the compositional phase diagram described by Fernandes \emph{et al}\cite{Fernandes2014}. The character of the phase transitions and their evolution with applied pressure, determined from our measurements, are consistent with this diagram, with the only difference being the observed initial decrease of $T_{\rm{s}}$ as pressure is lowered, rather than the displayed decrease. One must, however, consider the relative strengths of the interactions in such models\cite{Fernandes2014,Chubukov2015} and how these interactions evolve with pressure.  In particular, the stiffening of the elastic constants with increasing pressure can result in a decrease in $T_{\rm{s}}$ relative to its value at ambient pressure.  Similarly, an increase in the magnetic interaction, indicated by the increase in the magnetic ordering temperature and ordered magnetic moment with increasing pressure, can push $T_{\rm{s,~m}}$ for the coupled magneto-structural transition to higher temperature.
 
\begin{methods}
HE-XRD measurements were performed on the six-circle diffractometer at station 6-ID-D at the Advanced Photon Source, using 100.3 keV x-rays and a beam size of 100$\times$100 $\mu$m$^2$.  A sample with dimensions 120$\times$120$\times$20 $\mu$m$^3$ was loaded into a membrane-driven copper-beryllium diamond anvil cell (DAC).  A tungsten gasket with an initial thickness of 120 $\mu$m was pre-indented to a thickness of 70 $\mu$m, and a 660 $\mu$m hole was laser-drilled to accommodate the sample and pressure calibrants (ruby spheres and silver foil), as shown in Fig. 1e.  Helium gas was used as the pressure transmitting medium and loaded at $p$ = 0.5 GPa. The pressure was initially determined by the fluorescence lines from ruby spheres at ambient temperature, and, during the diffraction measurements, was determined $in$-$situ$ by analyzing selected Bragg peaks from the silver foil.  The DAC was mounted on the cold finger of a He closed-cycle refrigerator and temperature-dependent measurements were performed between $T$ = 5 and 300 K for various pressures.  The pressure was always changed at temperatures well above 120 K.  Extended regions of selected reciprocal lattice planes and the powder diffraction pattern of silver were recorded by a MAR345 image plate system positioned 1.474 m behind the DAC, as the DAC was rocked by up to $\pm$3.2$^{\circ}$ about two independent axes perpendicular to the incident x-ray beam.  High-resolution diffraction patterns of selected Bragg reflections were also recorded by employing a Pixirad-1 detector positioned 1.397 m behind the DAC while rocking around one of the two axes perpendicular to the x-ray beam.  The images in Fig, 3 show typical examples of diffraction patterns measured using the Pixirad-1 detector, and demonstrate the excellent mosaic of the single crystal under applied pressure, as evidenced by the well-split pattern of the ($HH0$) Bragg peaks (in tetragonal notation) due to the orthorhombic distortion\cite{Tanatar2009}. The orthorhombic lattice parameters were determined by fitting the Bragg peak positions after integrating the data over the transverse scattering directions.  This procedure was used for both the data recorded by the Pixirad-1 detector and the data recorded by the MAR345 image plate system.

In NFS, highly monochromatic synchrotron radiation from an electron bunch excites the $^{57}$Fe nuclei and the decay curve is measured as a function of time.  In the presence of a hyperfine interaction that splits the $^{57}$Fe energy level, oscillations in the scattered intensity with time (quantum beats) are observed and can be directly compared with conventional M\"{o}ssbauer spectroscopy\cite{Gerdau1999}. NFS spectra were collected at beamline 16-ID-D at the Advanced Photon Source with an incident energy monochromated to the $^{57}$Fe nuclear resonance at 14.4125 keV, with a resolution of 2 meV, and a cross-section of 35$\times$50 $\mu$m$^2$.  A $^{57}$FeSe single crystal of dimensions 50$\times$50$\times$18 $\mu$m$^3$ was loaded in to a membrane driven copper-beryllium DAC with 600 $\mu$m culet anvils, which allowed us to collect NFS data at $p$ = 2.5 GPa and 4.0 GPa. A non-magnetic Cu-Be gasket was pre-indented to ~55 $\mu$m, and a hole of diameter 270 $\mu$m was laser-drilled to accommodate the sample and pressure calibrant (ruby spheres).  Helium was loaded as the pressure-transmitting medium to enable hydrostatic pressure conditions. The DAC was mounted on the cold finger of a helium-flow cryostat, which achieved temperatures down to $T$ = 11 K. The intensity of the resonantly scattered photons in the forward direction, with 153.4 ns separation between the individual bunches, were recorded by an Avalanche Photo Diode detector. The program CONUSS\cite{Sturhahn2000}  was used to analyze the spectra and to determine the magnitude of the hyperfine field.  The NFS spectrum at ambient temperature was well modeled without a magnetic hyperfine field, fitting only the effective sample thickness and quadrupolar splitting, which yielded a value of 0.20(5) mm/s at $p$ = 0.8 GPa consistent with earlier reports\cite{Medvedev2009}.  The low-temperature spectra were satisfactorily fit by including a hyperfine magnetic field.  The two angles defining the direction of hyperfine magnetic field with respect to the incident beam were also determined in addition to the (slightly pressure and temperature dependent) effective thickness of the sample.  The values for the quadrupolar splitting were fixed to 0.15 and 0.20 mm/s at $p$ = 2.5 GPa and 4.0 GPa, respectively.
\end{methods}

\bibliography{FeSe}

\begin{addendum}
 \item The authors would like to acknowledge the assistance of D. S. Robinson, C. Benson, S. Tkachev, S. G. Sinogeikin and M. Baldini, and helpful discussions with R. J. McQueeney and R. M. Fernandes.  Work at the Ames Laboratory was supported by the Department of Energy, Basic Energy Sciences, Division of Materials Sciences \& Engineering, under Contract No. DE-AC02-07CH11358.  This research used resources of the Advanced Photon Source, a U.S. Department of Energy (DOE) Office of Science User Facility operated for the DOE Office of Science by Argonne National Laboratory under Contract No. DE-AC02-06CH11357. HPCAT operations are supported by DOE-NNSA under Award No. DE-NA0001974 and DOE-BES under Award No. DE-FG02-99ER45775, with partial instrumentation funding by NSF. Use of the COMPRES-GSECARS gas loading system was supported by COMPRES under NSF Cooperative Agreement EAR 11-57758 and by GSECARS through NSF grant EAR-1128799 and DOE grant DE-FG02-94ER14466. AEB acknowledges support from the Helmholtz Association via PD-226.

\textbf{Author contributions} *KK and *AEB contributed equally to this work. KK, AEB, SLB, PCC, AK, and AIG designed the measurements; AEB and VT grew the samples; KK, AEB, WTJ, BGU, PD, AS, DSR, and AK performed and analyzed the HE-XRD measurements; KK, AEB, WTJ, YX, EA, and AK performed and analyzed the NFS measurements; AIG, AK, AEB, KK, and PCC drafted the manuscript and all authors participated in the writing and review of the final draft.

\item[Competing Interests] The authors declare that they have no
competing financial interests.
 \item[Correspondence] Correspondence and requests for materials should be addressed to A. I. Goldman or A. Kreyssig~(email: goldman@ameslab.gov, kreyssig@ameslab.gov).
\end{addendum}

\newpage

\noindent {\bf Figure 1. Synchrotron high-energy x-ray diffraction and time-domain $^{57}$Fe M\"{o}ssbauer spectroscopy of FeSe under pressure.} Panels \textbf{a,} - \textbf{d,} show the evolution of the in-plane lattice parameters at various pressures determined from the splitting of the tetragonal ($HH0$) Bragg peaks. The color corresponds to detector intensities integrated over the transverse scattering directions.  Panel \textbf{e,} shows a photograph of the contents of diamond anvil pressure cell used for these measurements including the FeSe single crystal, and ruby and silver pressure indicators. Panels \textbf{f,} and \textbf{g,} display the time-domain $^{57}$Fe M\"{o}ssbauer spectra at 2.5 and 4 GPa, respectively, with these data sets vertically offset for clarity.  Grey lines are fits to the data described in the Methods Section.

\noindent {\bf Figure 2. Structural and magnetic order parameters of FeSe under pressure.} Panels \textbf{a,} and \textbf{c,} plot the orthorhombicity, $\delta(T)=(a_{\rm{OR}}-b_{\rm{OR}})/(a_{\rm{OR}}+b_{\rm{OR}})$, as a function of temperature at various pressures. Error bars indicate $2\sigma$, the fitting error. The inset in panel \textbf{a,} show a magnified view of the data close to the discontinuous change in $\delta$ near 19 K. A coexistence region for the orthorhombic and tetragonal phases is observed at 1.7 GPa and 3.1 GPa as indicated by open symbols and vertical lines. The inset of panel \textbf{c,} shows the relative integrated intensities of the tetragonal (T) and orthorhombic (OR) phases on warming and cooling respectively, at 1.7 GPa. Any thermal hysteresis is smaller than the point spacing (0.2 K), whereas the coexistence region spans 1.5 K. Panel \textbf{b,} displays the magnetic hyperfine field  $H_{\mathrm{hf}}$ derived from CONUSS fits\cite{Sturhahn2000} of the NFS data in Figure \textbf{1f,} and \textbf{g,}.

\noindent {\bf Figure 3. The Pressure-Temperature phase diagram of FeSe labeled with the orthorhombic (OR), magnetic (M) and superconducting (SC) ordered states.} The transition temperatures obtained from the present single-crystal HE-XRD measurements (red symbols) and fits of the NFS data (blue symbols). The grey erect triangles denote transition temperatures inferred from the resistivity measurements\cite{Kaluarachchi2016} on samples from the same batch used for the diffraction measurements and grey inverted triangles denote the measured values of $T_{\rm{c}}$ from previous work\cite{Miyoshi2014}. Thick lines in the Figure represent first-order phase transitions and thin lines correspond to second-order phase transitions. The dashed blue line shows a tentative extrapolation of these phase lines. The insets to the Figure show representative two-dimensional diffraction data in the respective $p$-$T$ region demonstrating the splitting of the tetragonal (660) Bragg peak in the orthorhombic phase.

\noindent {\bf Figure 4. Evolution of the character of the magnetic and nematic transitions in the spin-driven nematic theory\cite{Fernandes2014}.} For both panels the blue lines represent magnetic transitions, the red lines denote structural transitions, and the violet lines represents the joint magneto-structural transition.  The grey lines show the trends noted in theory\cite{Fernandes2014} which differ from those found in experiments. Thick lines denote first-order transitions, whereas the thinner lines denote second-order transitions. For panel \textbf{a,} the control parameter is partial element substitution (doping) for the case of BaFe$_2$As$_2$. For panel \textbf{b,} the control parameter is pressure for the case of FeSe. $T_{\rm{s}}$, $T_{\rm{s^*,m}}$ and $T_{\rm{s,m}}$ denote the structural transition temperature, the magnetic ordering temperature associated with the discontinuous change in the orthorhombicity, and the joint structural and magnetic transition temperature, respectively.

\newpage

\begin{center}
	\includegraphics[width=17cm]{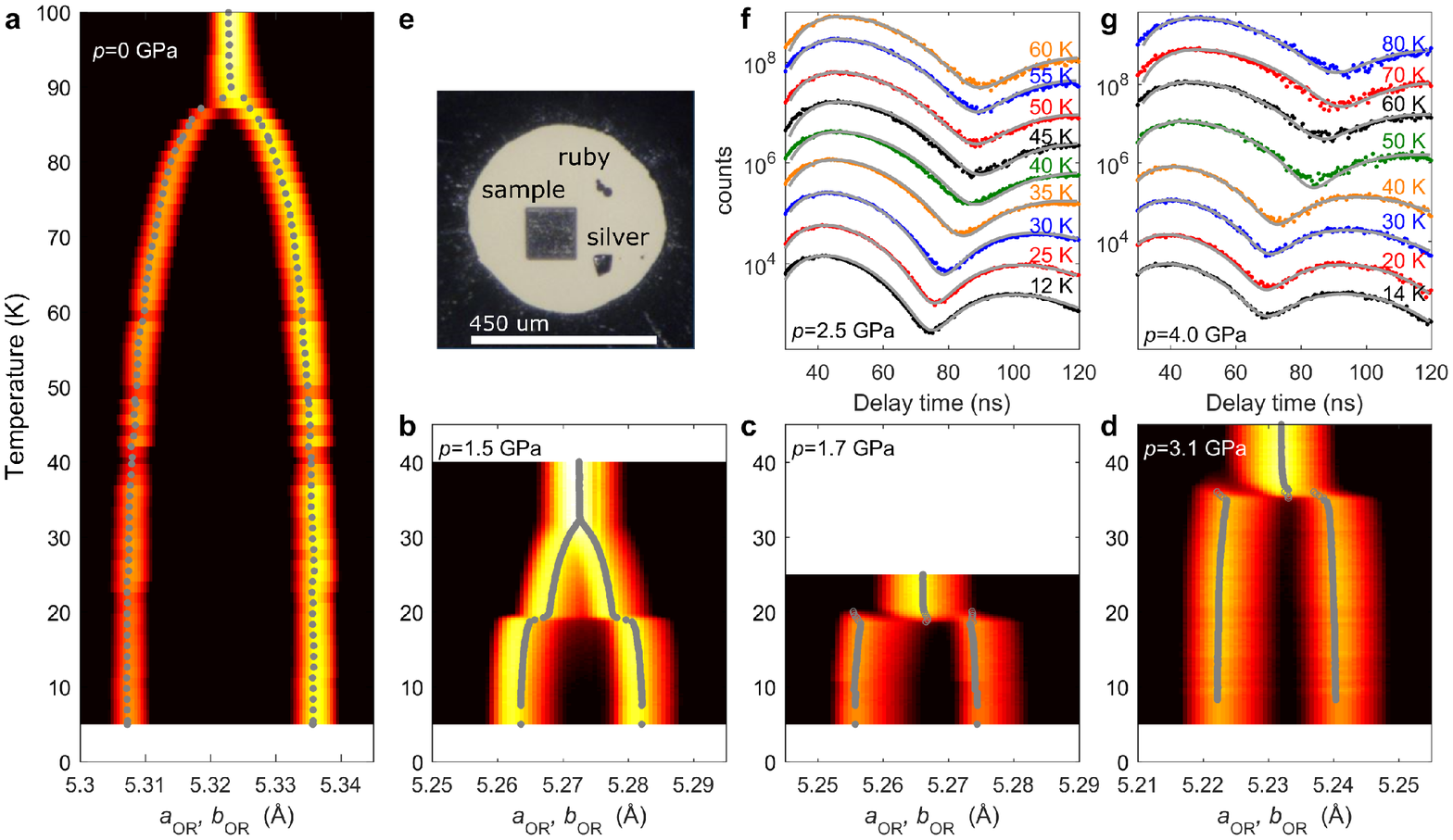}
\end{center}
\begin{center}Figure 1\end{center}

\newpage

\begin{center}
	\includegraphics[width=16cm]{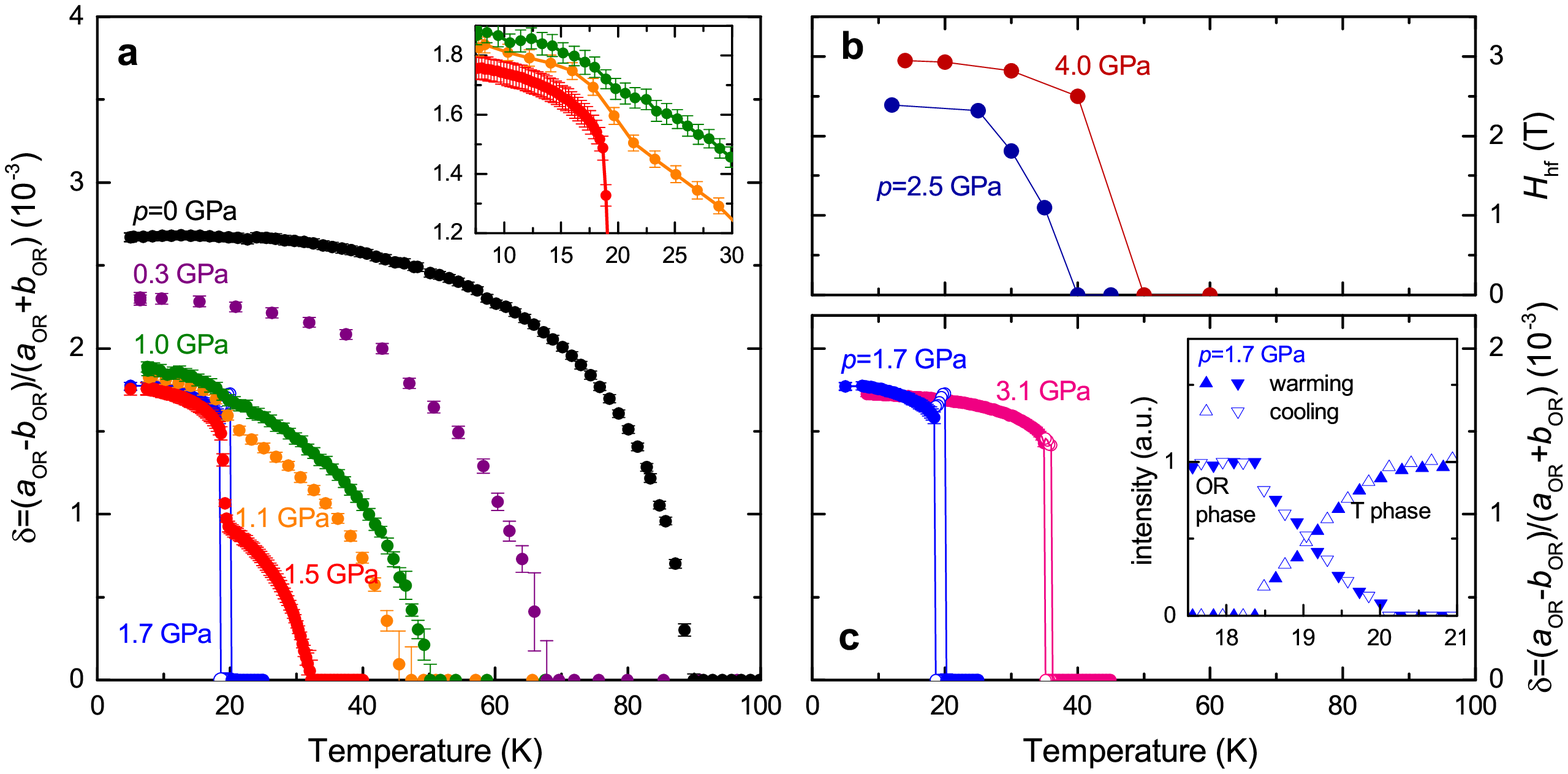}
\end{center}
\begin{center}Figure 2\end{center}

\newpage

\begin{center}
	\includegraphics[width=13cm]{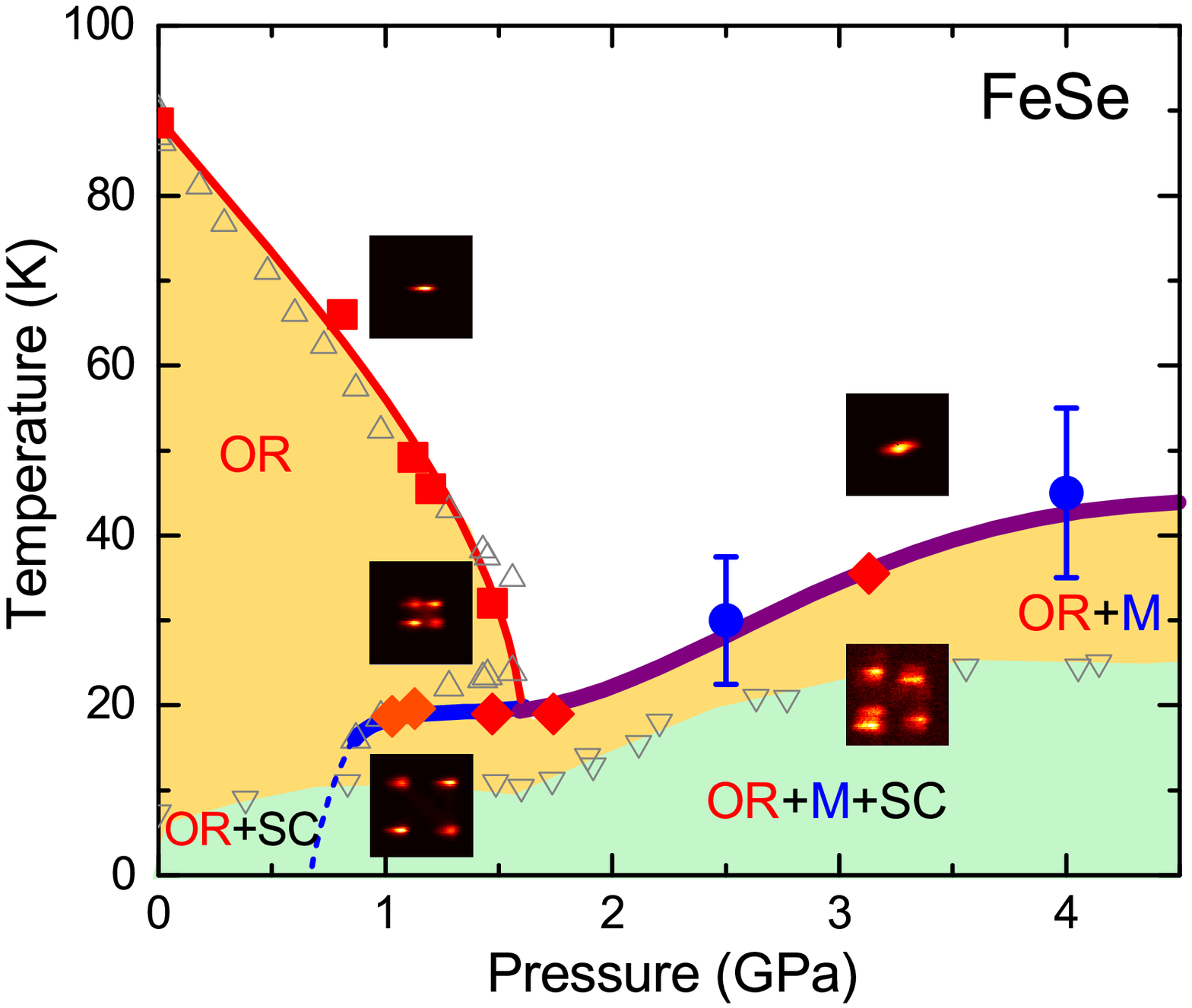}
\end{center}
\begin{center}Figure 3\end{center}

\newpage

\begin{center}
	\includegraphics[width=16cm]{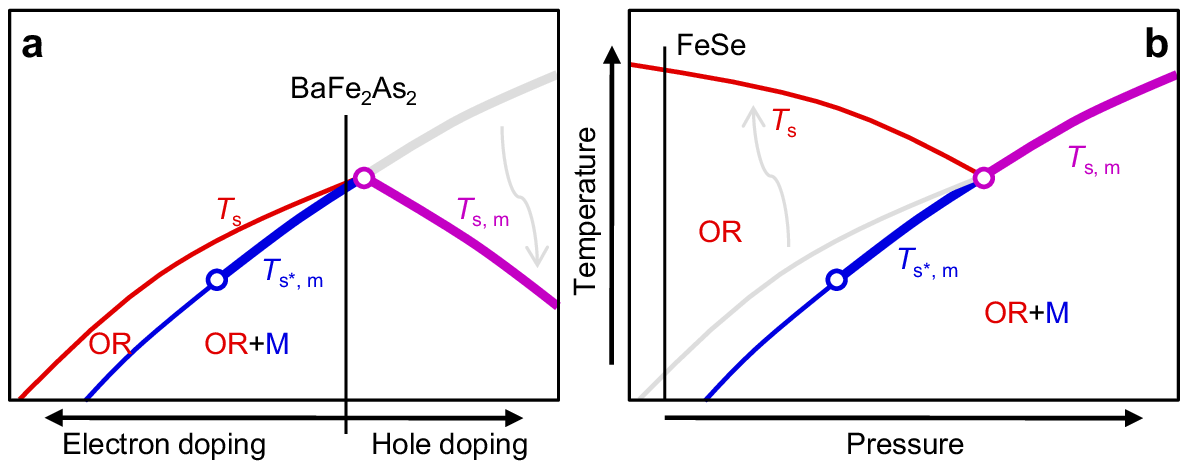}
\end{center}
\begin{center}Figure 4\end{center}

\end{document}